\documentclass[twocolumn,showpacs,preprintnumbers,amsmath,amssymb]{revtex4}

\usepackage{graphicx}
\usepackage{dcolumn}
\usepackage{bm}
\begin{document}


\title{Infrared scattering rate of overdoped Tl$_{2}$Ba$_{2}$CuO$_{6+\delta}$}

\author{Y. C. Ma}
\author{N. L. Wang}
\altaffiliation[Email: ]{nlwang@aphy.iphy.ac.cn}
\affiliation{%
Beijing National Laboratory for Condensed Matter Physics,
Institute of Physics, Chinese Academy of Sciences, Beijing 100080,
People's Republic of China
}%

\date{November 28, 2005}

\begin{abstract}
We present in-plane optical study on
Tl$_{2}$Ba$_{2}$CuO$_{6+\delta}$ single crystals with
substantially different $T_c$. The study reveals that the
overdoping does not lead to a further increase of the carrier
density but a decrease of scattering rate. The most significant
change occurs at low temperature and in the low frequency. A
characteristic spectral feature, seen most clearly for the
optimally doped sample and commonly ascribed to the mode coupling
effect, weakens with doping and disappears in the heavily
overdoped sample. Meanwhile, the optical scattering rate evolves
from a linear-$\omega$ dependence to an upward curvature
lineshape. Both the temperature and frequency dependence of the
scattering rate can be described by a power law relation. We
elaborate that the overall decrease of the optical scattering rate
originates from the increase of both the quasiparticle life time
and the Fermi velocity near the ($\pi$,0) region in the Fermi
surface.
\end{abstract}

\maketitle


For high-temperature superconductors (HTSC) and other strongly
correlated electron systems, the scattering rate of carriers
contains all relevant information of carrier interactions with
other quasi-particles or excitations. Investigating the frequency-
and temperature-dependent scattering rates in different regimes in
the phase diagram of HTSC is crucial for understanding the charge
behavior and dynamics in cuprate superconductors\cite{Basov}. In
the heavily underdoped regime, a non-monotonic form of the optical
scattering rate 1/$\tau(\omega)$ was observed in all crystals with
$y$$<$6.5 in YBa$_{2}$Cu$_{3}$O$_{y}$. A two-component model
offers a sufficiently accurate interpretation for this
behavior\cite{Lee,Basov}. As doping level increases up to nearly
optimal doping level, a linear frequency-dependent
1/$\tau(\omega)$ is commonly observed in the normal state which
could be described by a marginal Fermi-liquid (MFL) picture,
instead of the above two-component approach. Below 700 cm$^{-1}$
there is a sharp depression followed by an overshoot in
1/$\tau(\omega)$ in the superconducting state. Though not fully
understood, this structure has been generally attributed to the
interaction of the charge carriers with a bosonic
mode\cite{Hwang1,Basov}. With further doping, the cuprates go into
the overdoped regime. Recent studies indicate that the overall
scattering rate decreases gradually in the normal state.
Meanwhile, the above bosonic mode becomes more subtle and almost
disappears\cite{Hwang1,Basov}. Whether or not this mode is
responsible for the pairing of superconductivity remains a
challenging issue.

It is noted that the La-based cuprate (LSCO) could be doped
through the whole doping region\cite{Uchida}. However, this system
has relatively low superconducting transition temperatures with
maximum $T_c$$\sim$30 K. For Y- or Bi-based families with maximum
$T_c$$\sim$ 90 K, only limited range in the overdoping side could
be achieved in the phase diagram. Extensive experiments have been
carried out in the underdoped and optimally doped regimes,
comparatively, much less research work have been done in the
overdoped, especially in the heavily overdoped region due to the
difficulty of making overdoped samples, although there is no
complexity from the pseudogap issue. Among the systems with
maximum $T_c$$\sim$ 90 K, Tl$_{2}$Ba$_{2}$CuO$_{6+\delta}$
(Tl-2201) is rather exceptional. The system can be synthesized in
the whole region of the overdoped side with $T_{c}$ ranging from
90 K to 0 K\cite{Kubo,Shimakawa}. In comparison with the Bi-2201
or LSCO, the same overdoping level for Tl-2201 results in much
more suppression of $T_{c}$. In addition, Tl-2201 has a
well-ordered crystal structure with very flat CuO$_{2}$ layers far
apart from each other, about 11.6 \AA, while for Bi-2212, the
distance between the two CuO$_{2}$ planes in a unit cell is only
3.2 \AA, the bilayer splitting will inevitably
appear\cite{Gromko,Feng}. Many complications which could be found
in other cuprates such as in Bi-2212 or YBCO were not important
here. Therefore, Tl-2201 system is an ideal candidate for studying
the doping-dependent properties in the overdoped side of the
cuprate phase diagram. Recently, heat transport\cite{Proust},
inelastic neutron scattering (INS) experiments\cite{He}, polar
angular magnetoresistance oscillations (AMRO)\cite{Hussey} and
angle resolved photoemission spectroscopy (ARPES)\cite{Plate} have
been done on Tl-2201. Summarizing and analyzing the overall
results and discussions from various experiments on Tl-2201 will
help us to understand the HTSC in cuprates.

Among all the powerful tools for studying cuprates, infrared
reflectance spectroscopy probes the bulk properties and the
electron dynamics as the frequency-dependent dielectric constants
can be deduced from the reflectivity, therefore the spectrum of
electronic excitations in the energy range characteristic for
mobile carriers could be obtained. Furthermore, it would be much
helpful to understand the carriers behaviors in HTSC if ARPES
results were combined with the optical spectroscopy. Very
recently, we have successfully grown a series of Tl-2201 single
crystals with substantially different $T_{c}$. This provides us a
good opportunity to investigate the evolution of optical response
in the overdoped side. In this study, we concentrate on the charge
dynamics as revealed by the optical scattering rate.

Three Tl-2201 single crystals have been selected for optical
measurement, one nearly optimal doping with $T_{c}$=89 K, the
second one the mediately overdoped with $T_{c}$=70 K, and the
third the heavily overdoped sample with $T_{c}$=15 K. The widths
of the superconducting transitions are roughly 5 K for all the
three samples. All the single crystals have shiny planes with good
orientations and about 1mm$\times$1mm sizes, the first two samples
have been successfully grown by the flux method, see ref\cite{Ma},
the third one were also obtained by the same method but with
larger oxygen flux. The $T_{c}$s of the three samples were
characterized by the temperature-dependent dc resistivity. In our
experiments, we also measured the dc resistivity of one sample
with $T_{c}$=59 K, however, due to the small area, it was not
appropriate for infrared measurements.

\begin{figure}[t]
\includegraphics[width=0.85\linewidth]{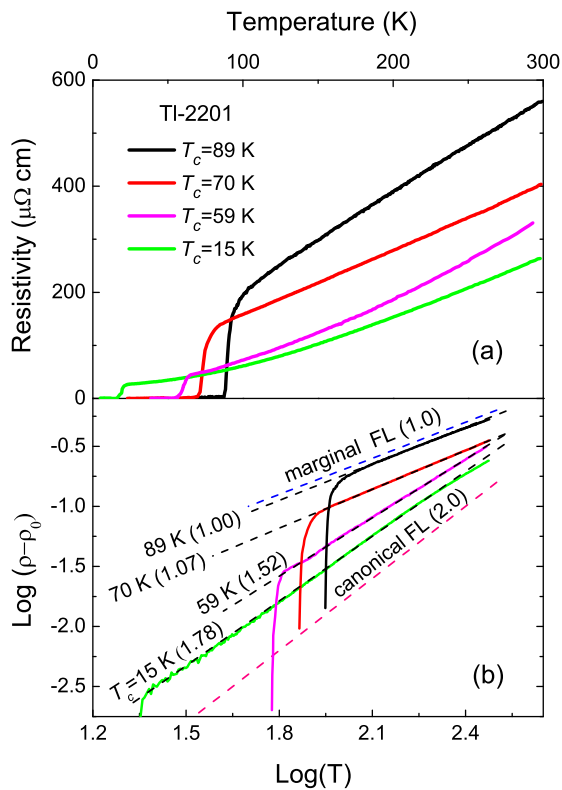}
\caption{\label{fig:epsart} (Color online) (a) The
temperature-dependent resistivity of Tl-2201 single crystals with
$T_{c}$=89 K, 70 K, 59 K and 15 K. (b) The log-log plot for
Tl-2201 single crystal corresponding to (a). The power law
behavior was seen in the whole measured temperature range. The
fitting lines with fixed power index were drawn as dashed lines.
The model for the MFL and the canonical Fermi liquid are included
for comparison.}
\end{figure}

The reflectance measurements from 100 to 22 000 cm$^{-1}$ for
E$\parallel$ab plane were carried out on a Bruker 66v/S
spectrometer. The samples were mounted on optically black cones in
a He-gas flowing cryostat with the experimental temperature range
8 K$\sim$320 K and temperature control better than 0.2 K. An
\textit{in situ} gold overcoating technique was used for the
experiment. The optical conductivity spectra were derived from the
Kramers-Kronig transformation. Hagen-Rubens relation was assumed
for the low-frequency extrapolation. At high frequency side, a
constant extrapolation was adopted up to 100 000 cm$^{-1}$, then a
R($\omega$)$\thicksim$ $\omega$$^{-4}$ relation was used.

\begin{figure}[t]
\includegraphics[width=0.80\linewidth]{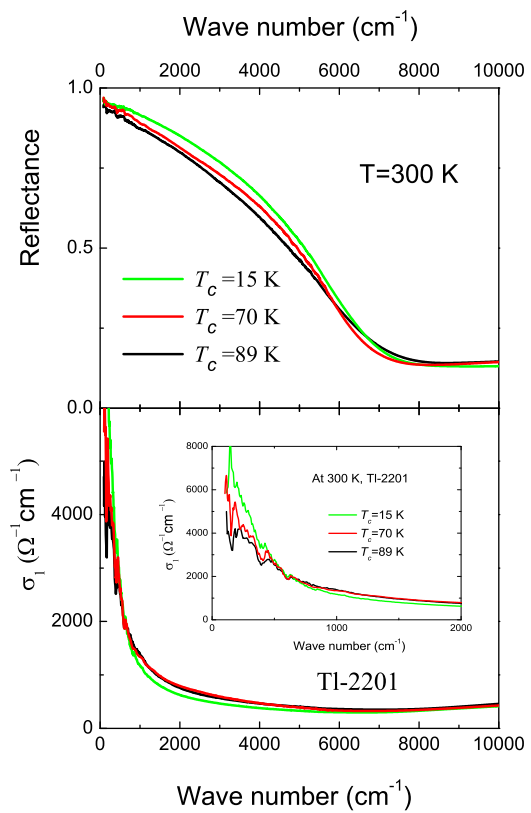}
\caption{\label{fig:epsart} (Color online) (a) The reflectance
data up to 10000 cm$^{-1}$ at room temperature of three Tl-2201
single crystals with the $T_{c}$=89 K, 70 K and 15 K,
respectively. (b) The corresponding conductivity at room
temperature calculated from the Kramers-Kronig transformation.
Inset: the conductivity spectra in the expanded region at low
frequencies.}
\end{figure}

The temperature-dependent dc resistivity ($\rho$($T$)) for Tl-2201
crystals is shown in Fig. 1(a). The resistivity decreases as
doping increases in the normal state. The sample with $T_c$=89 K
shows a linear $T$-dependent resistivity. As the samples become
overdoped, a superlinear $\rho$ vs $T$ relation appears. It is
found that the $T$-dependent resistivity actually follows a
power-law behavior $\rho(T)$$\sim$$T^{p}$, or equivalently, in the
log($\rho$-$\rho_0$)$\sim$log($T$) plot, it displays a linearity
behavior, where the slope is the power parameter and $\rho_0$ the
residual resistivity, as seen in Fig. 1(b). As $T_{c}$ decreases,
the slope or $p$ increases from 1.0 to 1.78. The $\rho$($T$)
curves are consistent with the reported data\cite{Kubo,Shimakawa}.

A comparison of the in-plane optical spectra of three Tl-2201
single crystals at room temperature is shown in Fig. 2. Despite of
the fairly large decrease in $T_c$ due to the overdoping, the
reflectance R($\omega$) is found to increase only slightly below
roughly 6000 cm$^{-1}$. This is very much different from the
spectral evolution in the underdoped side, where the R($\omega$)
below the reflectance edge changes dramatically with $T_c$ (or
doping level)\cite{Wang1}. One can estimate the effective carrier
number $N_{eff}$ below $\omega$ from the conductivity spectral
weight (SW) in terms of the partial sum rule:
\begin{eqnarray}
N_{eff}(\omega)=\frac{2mV_{cell}}{{\pi}e^2}SW(\omega)=\frac{2mV_{cell}}{{\pi}e^2}
\int_{0}^{\omega}{\sigma_1(\omega')d\omega'},
\end{eqnarray}
where $V_{cell}$ is a unit cell volume. In optical conductivity
spectra shown in fig. 2(b), only the WS below 600 cm$^{-1}$
increases for samples with lower $T_c$, the SW between 600-8000
cm$^{-1}$ decreases slightly. The overall spectral weight keeps
almost constant for the three samples, as seen in Table I. Those
results suggest that the effective carrier density does not
increase further with doping in the overdoped region, the major
change is the narrowing of low-$\omega$ Drude-like peak. This
narrowing originates from the reduction of the scattering rate,
which we shall address below. We note that such characteristic
spectral evolution is consistent with an earlier study on Tl-based
systems\cite{Puchkov1}.

\begin{table}
\caption{\label{tab:table2} The spectral weight distribution
calculated from Eq. (1) for three Tl-2201 crystals. All the data
has been normalized to SW(8000) of the optimally doped sample.}
\begin{tabular}{cccccc}
\hline%
    Sample      &$T_{c}$ (K)   &SW(600)   &SW(8000)-SW(600)  &SW(8000)\\
\hline
      A           & 89   & 0.312 & 0.688 &1.000 \\
      B           & 70   & 0.367 & 0.642 &1.019 \\
      C           & 15   & 0.425 & 0.582 &1.007 \\
\hline
\end{tabular}
\end{table}

Although the overall spectral weight does not change very much,
the low-$\omega$ R($\omega$) at low T changes significantly with
overdoping. Figure 3 shows the $T$-dependent R($\omega$) and
$\sigma_1$($\omega$) spectra from 100 to 2000 cm$^{-1}$ for the
three samples. For the $T_{c}$=89 K single crystal, the
R($\omega$) displays a well-known knee structure near 500
cm$^{-1}$, followed by a dip at higher energy close to 800
cm$^{-1}$ at $T$=10 K. This feature is commonly observed for
cuprate superconductors with high enough $T_c$. The characteristic
structure was usually ascribed to the coupling effect of electrons
with a bosonic
mode\cite{Basov,Hwang1,Carbotte,Abanov,Wang2,Tu,Wang3,Dordevic,Hwang2}.
Since such a lineshape causes the depression of low-$\omega$
conductivity leading to the missing of some spectral weight, the
shape is also related to the superconducting condensate. It is
seen clearly that the feature becomes weak as the samples become
overdoped, and disappears in the heavily overdoped $T_c$=15 K
sample. Note that for the heavily overdoped sample the
low-$\omega$ R($\omega$) is slightly depressed at low temperature
leading to a downward curvature. Such spectral lineshape was also
observed in earlier studies.\cite{Puchkov3,Puchkov2}

\begin{figure}
\includegraphics[width=1.0\linewidth]{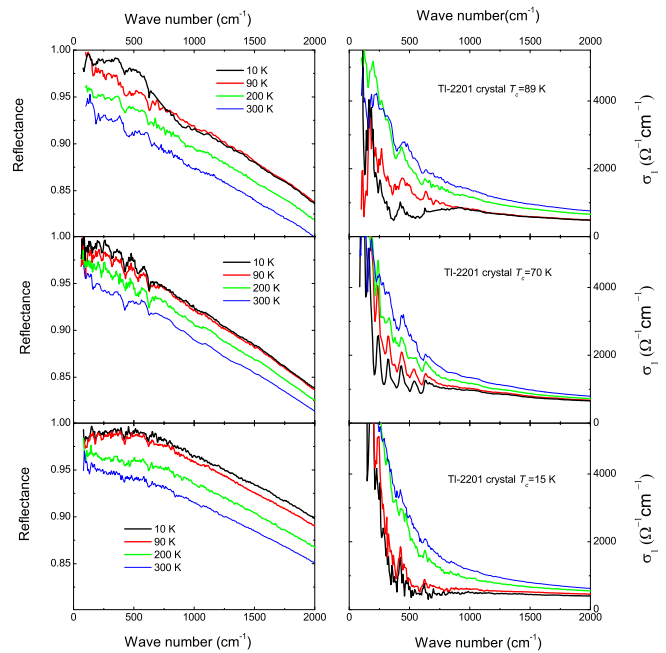}
\caption{\label{fig:epsart} (Color online) (a) The
temperature-dependent reflectance of three Tl-2201 single crystals
with the $T_{c}$=89 K, 70 K and 15 K from 100 to 2000 cm$^{-1}$.
(b) The corresponding conductivity spectra extracted from the
Kramers-Kronig transformation.}
\end{figure}

A useful way to analyze the carrier dynamics and the mode coupling
effect is in terms of the generalized Drude
model\cite{Hwang1,Puchkov2,Basov},
\begin{eqnarray}
\sigma(\omega)=
 \frac{i}{4\pi}
 \frac{\omega_p^2}{\omega-2\Sigma^{op}(\omega)},
\end{eqnarray}
where $\omega_p$ is the plasma frequency, which is related to the
N$_{eff}$ via the relationship
$\omega_p^2$=4${\pi}$$e^2$$N_{eff}$/m$V_{cell}$=8$\int_{0}^{\omega_c}\sigma_1(\omega')d\omega'$
after choosing a proper high-frequency limit $\omega_c$,
$\Sigma^{op}$ is so-called optical single-partical self-energy:
$\Sigma^{OP}$($\omega$)=
$\Sigma_1^{OP}$($\omega$)+$i$$\Sigma_2^{OP}$($\omega$). The
imaginary part determines the frequency-dependent carrier
scattering rate via
$\Sigma_2^{OP}$($\omega$)$\equiv$$-$1/2$\tau$($\omega$), while the
real part is related to the mass enhancement via
$\Sigma_1^{OP}$($\omega$)$\equiv$$\omega$(1$-$m*/m)/2.\cite{Hwang1}
In Fig. 4, we show the optical scattering rate as well as the real
part of the optical self-energy for the three samples at different
temperatures. Here, the plasma frequency
$\omega_p$=1.5$\times$10$^4$ cm$^{-1}$, determined by integrating
the conductivity up to $\omega_c$=1 eV, was used for all three
samples. We can see that, for nearly optimally doped sample at 10
K, the scattering rate shows an onset increase near 500 cm$^{-1}$
and an overshoot at frequency corresponding to the dip in the
low-$T$ R($\omega$). In the real part of the self-energy a sharp
peak was seen in the low-$T$ curve. The optical resonance
structure was widely attributed to the coupling to a bosonic mode.
Clearly, this feature weakens with doping and vanishes for the
heavily overdoped sample with $T_c$=15 K.

\begin{figure}[t]
\includegraphics[width=1.0\linewidth]{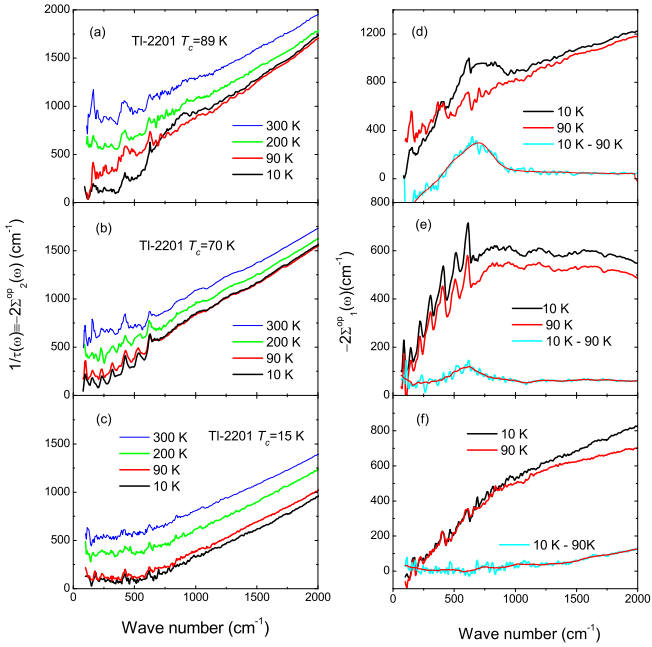}
\caption{\label{fig:epsart} (Color online) a-c: the doping- and
temperature-dependent optical scattering rate 1/$\tau$($\omega$)
from 100 to 2000 cm$^{-1}$. d-f: the real part of the optical
self-energy of Tl-2201 in the normal state (T=90 K) and
superconducting state (T=10 K), as well as the corresponding
differences between them. The corresponding smoothed differences
have also been shown. (a, d), $T_{c}$$\sim$89 K (nearly optimal
doped); (b, e), $T_{c}$$\sim$70 K (mediately overdoped); (c, f),
$T_{c}$$\sim$15 K (heavily overdoped).
 }
\end{figure}

Hwang $et$ $al.$ studied the self-energy effects in optics for
different dopings and temperatures in Bi-2212 system\cite{Hwang1}.
They identified that the peak in the real part of the optical
self-energy is closely related to the ARPES self-energy obtained
by Johnson et al. performed at the nodal point\cite{Johnson}. They
also found that as the doping level increases, the contribution of
the bosonic resonance to the self-energy weakens and disappears in
the highly overdoped regime. Similar trend of ARPES self-energy
effects with doping near the antinodal point ($\pi$,0) was also
observed by Kim \emph{et al.} on Bi-2212 system\cite{Kim}.
However, it deserves to remark that both Hwang \emph{et al.} and
Kim \emph{et al.} extrapolated their data to a critical doping
level of $\delta_c$=0.23$\sim$0.24 where they claimed that the
anomalous peak in the real part of the self energy would no longer
exist, but they did not have real data for such doping levels. In
our experiments, the corresponding doping level of the
$T_{c}$$\sim$15 K overdoped Tl-2201 sample is about
$\delta\simeq$0.26 based on the universal relation between T$_c$
and the doping level for HTSC\cite{Presland}. So it offers an
opportunity to clarify the issue. Indeed, the feature caused by
the narrow mode is not visible for the heavily overdoped sample.
The consistency of the spectral evolution in two different systems
suggests that this is the generic property of the Cu-O layers in
the overdoped regime of HTSC.

The nature of the mode involved in the coupling remains under
intense debate. Here we remark that the observed spectral feature
is not consistent with a coupling with a phonon, but could still
be reconciled with a coupling to magnetic excitations. According
to neutron scattering experiments, the magnetic resonance mode at
($\pi, \pi$) shows a strong doping dependence: (1) The resonance
has the highest energy E$_r$ at the optimal doping, but shifts to
lower energy with either increasing or decreasing the doping
levels with a preserved E$_r$/$k_B$$T_c$ ratio. For Tl-2201, the
resonance is seen at E$_r$=47 meV at optimal doping, and the ratio
of E$_r$/$k_B$$T_c$$\simeq$6.\cite{He,Sidis,Bourges} (2) The
spectral weight of the resonance drops sharply in the overdoped
regime at the doping level close to
$\delta\sim$0.19.\cite{Pailhes} Apparently, in the case of
coupling to the magnetic resonance, the coupling feature should
also shift to lower frequency and weaken with overdoping.
Qualitatively this is indeed what we observed here, as well as the
data obtained by Hwang et al. for Bi-2212 system\cite{Hwang1}.
Additionally, the inhomogeneity, which is inevitably present in
those samples, would lead to a further damping of the mode effect.
Since neutron experiments were performed only in limited range in
the overdoped side, it is not clear whether or not the magnetic
resonance mode would disappear in the heavily overdoped sample,
then a precise and complete comparison could not be made. Another
possibility for the vanishing spectral feature in the heavily
overdoped T$_c$=15 K sample is that the feature shifts to the
region below our measurement frequency. It is well-known that the
characteristic mode-coupling feature in optics should appear at an
energy 2$\Delta$+E$_r$,\cite{Basov,Hwang1,Carbotte,Abanov} where
$\Delta$ stands for the maximum of the d-wave superconducting gap.
The values of the superconducting gap as a function of hole doping
for Tl-2201 are given by Hawthorn et al.\cite{Hawthorn}. For
T$_c$=15 K, the 2$\Delta\sim$5 meV, while the resonance is
expected to appear at E$_r$$\sim$~7.8 meV based on the above
scaling ratio between E$_r$ and $T_c$\cite{Sidis,Bourges}, then
the coupling feature in optics should appear at 100 cm$^{-1}$.
This is already out of the range for getting reliable data on such
small size samples in our measurement. Note that, in case of
coupling with a phonon at energy E$_p$, one can carry out the same
discussion, but in such a case, the characristic energy of the
phonon should remain almost constant as a function of hole doping:
E$_p$ should be about 40-50 meV. The mode coupling effect should
be observed around 320-400 cm$^{-1}$, in the heavily doped sample.
That is not consistent with the data presented here.

Besides the vanishing of mode structure, the differential spectra
between 10 K and 90 K of the real part of optical self-energy also
show interesting change with doping at high frequencies (800-2000
cm$^{-1}$). As seen from Fig. 4 (d-f), the signal is weak for the
nearly optimally doped sample, and decreases slightly with
increasing energy. The signal increases in the $T_c$=70 K sample
and is almost constant as a function of the energy. It further
increases and shifts to higher energy in the $T_c$=15 K sample.
Note that the evolution of the differential spectra is closely
related to the forms of the $T$-dependent reflectance R($\omega$),
as shown in the left panel of Fig. 3. For the nearly optimally
doped sample, R($\omega$) is almost temperature-independent at
high-$\omega$ below 90 K; however, in the frequency range of
750-1500 cm$^{-1}$, the reflectance at 10 K is slightly lower than
that at 90 K due to the effect of mode coupling or dip structure
appearing near 800 cm$^{-1}$. As the samples become overdoped, the
reflectance spectra below 90 K separate from each other gradually.
The spectral difference can be naturally ascribed to the different
$\omega$- and T-dependent carrier scattering rates for different
doping levels. In fact, the real part of optical self-energy is
linked with the optical scattering rate (i.e. imaginary part)
through the Kramers-Kronig relationship.

\begin{figure}[t]
\includegraphics[width=0.85\linewidth]{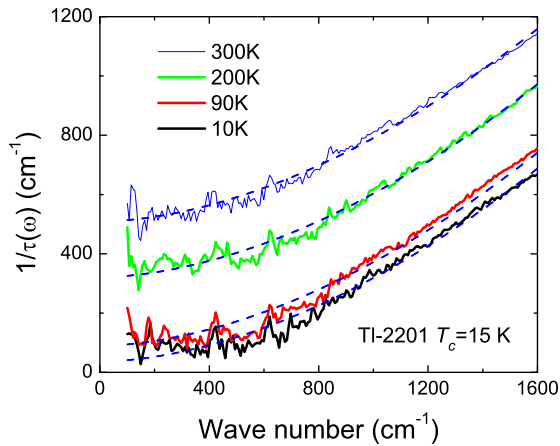}
\caption{\label{fig:epsart} (Color online) The scattering rates of
Tl-2201 with $T_{c}$$\sim$15 K from 100 to 1600 cm$^{-1}$ at
different temperatures and the corresponding fittings.}
\end{figure}

We note that, similar to the $T$-dependent dc resistivity, the
$\omega$-dependent behavior of the scattering rate also changes
with doping. For the $T_c$=89 K sample, the 1/$\tau$($\omega$)
displays a linear-$\omega$ dependence in the normal state. As the
sample becomes overdoped, the 1/$\tau$($\omega$) shows an upward
curvature. The behavior is seen very clearly in the heavily
overdoped sample with $T_c$=15 K. As the dc $\rho$($T$) follows a
power-law dependence, we expect the same power-law behavior for
the $\omega$-dependent scattering rate. Therefore, we suggest that
the scattering rate may follow approximately the relation:
\begin{eqnarray}
1/\tau(\omega,T)=
1/\tau_0+\alpha(k_{B}T)^{p}+\beta(\hbar\omega)^{p}.
\end{eqnarray}
Here, the first term in the right-hand side is from the impurity
scattering, $\alpha$ and $\beta$ are constants, $p$ ranges from 1
to 2. Indeed, we found that this formula successfully describes
both $T$- and $\omega$-dependent of the scattering rate. Figure 5
shows the 1/$\tau$($\omega$) data for the heavily overdoped sample
at different temperatures together with fitting curves using the
above formula up to 1600 cm$^{-1}$. Here, we use the value
$p$=1.78 determined by the $T$-dependent dc resistivity in our
fitting. We found that $\alpha$/$\beta$$\sim$$(1.6\pi)^{2}$ best
reproduces the 1/$\tau$($\omega$) results. Above 1600 cm$^{-1}$,
the 1/$\tau$($\omega$) tends to become $\omega$-linear dependence,
whereas at low-$\omega$, a small upturn of 1/$\tau$($\omega$) at
low-$T$ was seen, which corresponds to the unconventional small
depression in the reflectance spectrum as mentioned above. This
effect is commonly seen for the very overdoped cuprates and
addressed in earlier studies\cite{Puchkov3,Puchkov2}. It was
suggested to be related to the defects in the samples. The fitting
results indicate that the $T_c$=15 K sample is still at the
intermediate state from MFL to a canonical Fermi-liquid. It should
be mentioned that the canonical Fermi-liquid behavior ($p$=2)
considered here is for the case of the three-dimensional (3D)
electron system. For a 2D system, the scattering rate should
behave as 1/$\tau$$\sim$
$\alpha$$T^2$ln$T$+$\beta$$\omega^2$ln$\omega$.\cite{Zheng} The
fact that both the temperature-dependent dc resistivity and
optical response as a function of frequency follow the trend
towards a canonical Fermi liquid fairly well indicates indeed that
the system tends to become 3D-like in the overdoped regime,
similar to the results obtained from AMRO experiment by Hussey et
al.\cite{Hussey}. This behavior was also consistent with the
results of the $T$-dependent $\rho_c$/$\rho_{ab}$ in ref
\cite{Hermann}. Nevertheless, the real Fermi liquid state is
expected to show up only in non-superconducting samples.

\begin{figure}
\includegraphics[width=0.85\linewidth]{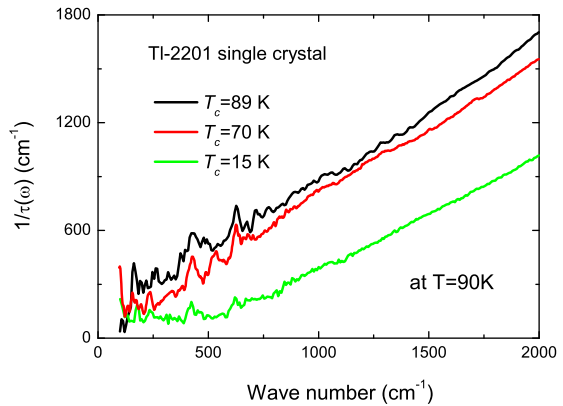}
\caption{\label{fig:epsart} (Color online) A comparison of the
scattering rates for three Tl-2201 single crystals from 100 to
2000 cm$^{-1}$ in the normal state at 90 K. }
\end{figure}

Another remarkable result is that the $\omega$-dependent
scattering rate decreases gradually with doping. Implication for
this effect was already seen from the Drude component narrowing of
the room-temperature conductivity spectra displayed in Fig. 2(b).
A careful comparison of the scattering rate spectra indicates that
the decrease is present at every measured temperature. As an
example, we show in Fig. 6 such a comparison at 90 K for the three
samples.

It should be noted that the quasiparticle lifetimes are highly
anisotropic around the Fermi surface. The optical scattering rate
deduced from the generalized Drude model, Eq. (2), can be taken as
an effective average of the scattering rates over Fermi surface,
although several comparative studies on ARPES and optical
spectroscopy revealed that the optical scattering rate is related
more closely to the imaginary part of the self-energy of
quasiparticle near ($\pi/2,
\pi/2$)\cite{Hwang1,Johnson,Basov2,Santander-Syro}. ARPES
experiments revealed that in underdoped curpates, the
quasiparticles are sharp near ($\pi/2, \pi/2$) and ill defined
around ($\pi$, 0) in the normal state. Upon increasing doping, the
antinodal quasiparticles sharpen up, but they remain broader than
the nodal quasiparticles all the way to optimal
doping\cite{Damascelli}. With further increasing doping in the
overdoped side, the antinodal quasiparticles become narrower than
the nodal quasiparticles\cite{Plate} (see the schematic picture of
Fig. 7). Those results indicate clearly that the major increase of
the quasiparticle life time (or reduction of scattering rate) is
in the antinodal region. Actually, the recent ARPES study on
overdoped Tl-2201 reveals that the nodal quasiparticle peak
becomes even broader in heavily overdoped sample than in
intermediate overdoped sample\cite{Plate}. Naively, one can
ascribe the reduction of the optical scattering rate when
overdoping to the increase of the lifetime of antinodal
quasiparticles on the basis of the ARPES results.

\begin{figure}
\includegraphics[width=0.95\linewidth]{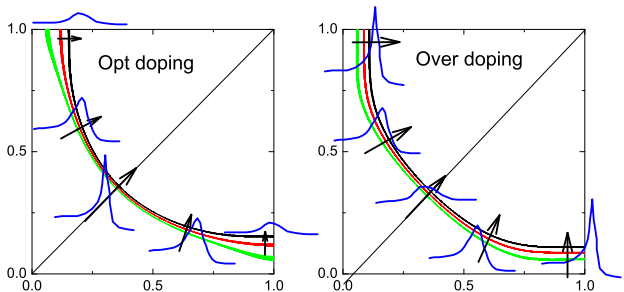}
\caption{\label{fig:epsart} (Color online)A schematic picture for
the Fermi surface, the quasiparticle peak width, the Fermi
velocity at different momentums at the optimal doping (left panel)
and overdoping (right panel). The blue curves show the Fermi
surfaces, and the grey curves are the equal energy contours. The
Fermi velocity (indicated by the red arrow) is the gradient of the
Fermi surface, which is apparently large at nodal direction and
small near ($\pi$, 0) or (0, $\pi$). With overdoping, the Fermi
velocity increases in the antinodal region. The black curves
indicate the quasiparticle peaks as seen in ARPES. The width
reflects the inverse of quasiparticle lifetime. For optimally
doped sample, the quasiparticle peak is sharp near the nodal point
but broad near ($\pi$, 0) or (0, $\pi$). In the very overdoped
sample, a reversal of those anisotropy is seen.\cite{Plate}}
\end{figure}

However, a more careful consideration suggests that the lifetime
increase of the antinodal quasiparticles is not the sole reason
for the reduction of the optical scattering rate derived from
extended Drude mode, the change of the Fermi velocity arising from
the gradual change of the shape of Fermi surface with doping,
especially near the antinodal region, also contribute to the
transport. Let us elaborate this point in more detail. According
to the generalized Drude model, Eq. (2), the optical scattering
rate is derived from the conductivity as
1/$\tau$($\omega$)=($\omega^2/4\pi$)Re(1/$\sigma$($\omega$)). In
the semiclassical approximation, the frequency-dependent
conductivity can be expressed as\cite{Ashcroft}:
\begin{eqnarray}
\sigma(\omega)\propto
\int{\frac{v_k^2}{\tau^{-1}(\varepsilon_k)-i\omega}} \left(
\frac{\delta f}{\delta\varepsilon}\right)
_{\varepsilon=\varepsilon_k}d^2k,
\end{eqnarray}
where $\textbf{v}_k$ and $\varepsilon_k$ are the carrier bare
(band-structural) velocity and energy, respectively,
$\tau^{-1}$($\varepsilon_k$) is the transport scattering rate
(which is apparently different from the optical scattering rate),
and $f$ is the Fermi distribution function. The derivative of $f$
is peaked at the Fermi energy. Actually, this equation was used by
Santander-Syro et al.\cite{Santander-Syro} to illustrate that the
in-plane conductivity was mostly sensitive to the nodal
quasiparticles. According to this equation, the Fermi velocity
$\textbf{v}_k$ appears as a weighted factor in the integration.
The larger the Fermi velocity and the smaller the scattering rate
(i.e. the longer quasiparticle lifetime), the larger contribution
it has to the transport. For underdoped or even optimally doped
cuprates, the Fermi surface has curvature near ($\pi$/2, $\pi$/2)
but flat near ($\pi$, 0) or (0, $\pi$), as shown in the left panel
of Fig. 7. Then, the gradient of the Fermi surface, which gives
the Fermi velocity
$\textbf{v}_k$=(1/$\hbar$)$\nabla_\textbf{k}\varepsilon(\textbf{k})$,
is highly anisotropic. Based on the equal energy contours shown as
the schematic picture in Fig. 7,\cite{Norman} the momentum change
for the same energy interval in the direction perpendicular to the
Fermi surface is larger near ($\pi$, 0) or (0, $\pi$) than near
($\pi$/2, $\pi$/2). This means smaller Fermi velocity near ($\pi$,
0) or (0, $\pi$) than near ($\pi$/2, $\pi$/2). As a result, the
in-plane transport is governed by the quasiparticles in the nodal
region. However, for the overdoped cuprates, the Fermi surface
becomes more circle-like (see the right panel of Fig. 7), the
anisotropy of $\textbf{v}_k$ resulting from the Fermi surface
gradient becomes much smaller, then the contribution from the
quasiparticles in the antinodal region will increase. To
summarize, both the increase of the Fermi velocity and the
reduction of carrier scattering near the antinodal region
contribute to the in-plane conductivity, and as a result, leading
to the reduction of the in-plane optical scattering rate as
defined from the generalized Drude model.

In summary, infrared studies of Tl-2201 single crystals for three
various doping levels have been carried out. Different from the
spectral evolution in the underdoped side, the overall spectral
weight does not increase further with doping in the overdoped
side, the major change is the narrowing of the Drude-like
component due to the reduction of the carrier scattering rate. On
the other hand, the low-temperature and low-frequency spectra show
significant change with doping. A clear knee structure followed by
a dip at higher frequency in the reflectance was observed for a
sample close to the optimal doping. Correspondingly, an onset
increase followed by an overshoot could be seen in optical
scattering rate spectra. Those features, which were commonly
ascribed to the mode coupling effect, weaken with doping and
vanish for the heavily overdoped sample. We point out that this is
a generic property for the cuprates in the overdoped side.
Furthermore, we found that the optical scattering rate evolves
from a linear-$\omega$ dependence near optimal doping to a shape
with upward curvature upon further doping. Both the temperature
and frequency dependence of the scattering rate can be described
by a power law relation. We also elaborate that the overall
decrease of the optical scattering rate originates from the
increase of both the quasiparticle life time and the Fermi
velocity near the ($\pi$,0) region in the Fermi surface.

\begin{acknowledgments}

This work was supported by the Ministry of Science and Technology
of China (973 project No. 2006CB601002), the National Science
Foundation of China, the Knowledge Innovation Project of Chinese
Academy of Sciences.
\end{acknowledgments}

\end{document}